# A *Bootstrap Theory*: the SEMAT Kernel Itself as *Runnable Software*

Iaakov Exman

Software Engineering Department

The Jerusalem College of Engineering - Azrieli

Jerusalem, Israel

iaakov@jce.ac.il

*Abstract—* The SEMAT kernel is a thoroughly thought generic framework for Software Engineering system development in practice. But one should be able to test its characteristics by means of a no less generic theory matching the SEMAT kernel. This paper claims that such a matching theory is attainable and describes its main principles. The conceptual starting point is the robustness of the Kernel alphas to variations in the nature of the software system, viz. to software automation, distribution and self-evolution. From these and from observed Kernel properties follows the proposed *bootstrap principle*: a software system theory should itself be a runnable software. Thus, the kernel alphas can be viewed as a top-level ontology, indeed the Essence of Software Engineering. Among the interesting consequences of this bootstrap theory, the observable system characteristics can now be formally tested. For instance, one can check the system completeness, viz. that software system modules fulfill each one of the system requirements.

## I. INTRODUCTION

The SEMAT kernel, recently published in the Kernel book [1], is coined the Essence of Software Engineering. The kernel characteristics were conceived and analyzed to serve as a generic framework to develop software systems in practice.

But one may ask:

- In which sense and to what extent does the kernel and its alphas, really represent the essence of software engineering?
- Are the kernel and its characteristics testable in a formal sense, without losing its generality?

It turns out that solving these issues is feasible. This paper offers a systematic approach to attain this goal. We assume familiarity of the reader with the SEMAT kernel.

Our investigation starts by testing the robustness of the Kernel alphas to variations of the software system nature. This is done by submitting them to thought experiments – *gedanken experiments* – in the tradition of physics and philosophy [2].

Concomitantly, we recognize that the SEMAT kernel has a hierarchical structure, with the kernel alphas in its top, and it is extensible towards its bottom like a software system.

Another observation is that the kernel is actionable, i.e. one can associate with it a behavior. Indeed in a practitioner's workshop on the SEMAT kernel this was concretely illustrated by "running" tangible cards of printed cardboard (see chapter 4 in the Kernel book [1]). One could actually use running software instead of cards.

This is the decisive step towards the *bootstrap principle* of this work: we claim that a software theory should itself be runnable software. Based upon this principle one can answer in a positive vein the questions raised above.

The remaining of the paper is organized as follows. First, it raises the issue whether the kernel alphas are robust enough to encompass software automation (section II), then it deals with distributed software systems (section III), next a third similar issue refers to self-evolving software systems (section IV). We formulate the *bootstrap principle* of this work (section V), and view the Kernel alphas as a top-level ontology (section VI). The paper is concluded with a discussion.

## II. TESTING THE SOFTWARE AUTOMATION HYPOTHESIS

Here we start to investigate, in which sense the Kernel Alphas and their consequences are actually the Essence of Software Engineering, i.e. are they robust enough to deal with any variation of software system? We begin with the issue of whether Kernel Alphas fit software automation.

### A. Alphas that fit Automation

The current Alphas' Kernel assumes that some of the Alphas – say stakeholders, team, work – involve or are performed by people. Citing the Kernel book definitions:

- **Stakeholders:** The *people*, groups, or organizations that affect or are affected by a software system;
- **Team**: The group of *people* actively engaged in the development, maintenance, delivery and support of a specific software system;
- **Work**: Activity involving *mental* or physical effort done in order to achieve a result.

Is the assumption that some of the Alphas involve people or people activities essential?

It seems that this assumption is easily modified to allow partial or complete automation of some or all the Alphas, without compromising the overall Alphas consistency. In other words, in some software systems, instead of people, some or all of the Alphas could themselves be software subsystems.

In this case one could have some of the *stakeholders* that are themselves software subsystems different from the Alpha "Software System". One could also have mixed *teams* in which certain team members are people and other team members are themselves software subsystems different from the Alpha "Software System" and at the same time different from the Stakeholders "Software subsystems". Then, *work* would involve not only mental or physical effort, but also some kind of software computational effort.

### B. Importance and Motivation for Automation

There are two deep reasons to justify software automation:

a. *Reliability* – Automation exists to increase reliability of expensive and critical systems;
b. *Formalization* – Automation probably facilitates formalization of any theory, even the sought General Theory of software engineering;

We first refer to automation for the sake of formalization. In order to better understand the behavior of a *software stakeholder* or a *software team* member one should model them. But software stakeholders will be made of software just as any other piece of software. Therefore their modeling is essentially like models of any kind of software. It should be easier to model – thus formalize – the software stakeholder than a human stakeholder. In other words, automation facilitates formalization of the theory.

*C. Automation for Reliability: a Gedanken Experiment*

Concerning automation for reliability, let us assume that a satellite – similar to Curiosity [3] – is sent to planet Mars. Such a satellite should be able of self-testing either periodically or in response to triggering events.

Suppose a failure/bug is found in a software sub-system of the satellite. There should be a decision mechanism that decides to send to the planet Earth station a signal requesting a replacement for the failed sub-system. Suppose further that this replacement does not exist and must be urgently developed to return the satellite to normal activity. The decision mechanism could be an a priori automatic system or a more complex mechanism that weighs alternatives. The latter mechanism certainly is a kind of stakeholder signaling an *opportunity*. This *software stakeholder* could maintain a dialogue with the human stakeholder in the planet Earth station, and so on. One could easily extend this idea to more complex software stakeholders. This scenario is not science fiction as satellite developers and software manufacturers actually send upgrades automatically to their customers [3].

The point here is that the same kernel alphas – with only slight revision to include software agents in their definitions – encompass software automation.

## III. TESTING THE DISTRIBUTED SOFTWARE HYPOTHESIS

We continue with the issue of whether Kernel Alphas fit distributed software systems.

*A. Alphas for Distributed Systems*

A further generalization of automated Alphas is software distribution. Assume now that instead of a single satellite one has a satellite group on Mars – in the Brooks' [4] robots style.

Not all of the satellites on Mars need to be in direct communication with the planet Earth station. They elect among themselves a sub-set of communicators. But once a failed sub-system is found in one of the satellites, two things happen: 1- the news are spread among the satellites; 2- they have a consultation among themselves whether a replacement is needed or not. So, we have a *distributed* group of *software stakeholders*, while only a sub-set is signaling to planet Earth.

An equivalent situation refers to *distributed teams* of software workers.

## IV. TESTING THE SELF-EVOLVING SOFTWARE HYPOTHESIS

A final example is the issue of whether Kernel Alphas fit self-evolving software systems.

*A. Alphas for Self-Evolving Systems*

Besides automation, we can think of software systems as having some characteristics of organisms – living or artificial.

Robots certainly fit the SEMAT Kernel definition of embedded software systems. Nowadays one has concrete examples of robots that compose themselves in real-time in order to overcome obstacles in their surroundings. This is an example of self-evolving systems, in which the surroundings provide the "*opportunity*" for development.

Self-stabilizing systems (see e.g. S. Dolev [5]) is a branch of computation originated by an idea of Dijkstra [6]. These software systems modify themselves to restore the desirable conditions of execution. Again, one can extend this kind of reasoning to include *stakeholders* and working *team* members as sub-systems of the software.

*B. Alphas are the Essence*

Extrapolating from the testing of the last three hypotheses – viz. automation, distribution and self-evolution – we may get to the gratifying conclusion that the Kernel Alphas, with only some mild modification/addition, indeed are a compact *Essence* and robust conceptualization of software systems.

## V. THE BOOTSTRAP PRINCIPLE: THEORY AS RUNNABLE SOFTWARE

Assuming that the Kernel Alphas are a solid basis for a general theory, we proceed to theory formulation.

### A. Similarity of the SEMAT Kernel to a Software System

We list here four arguments about the similarity of the SEMAT Kernel to a software system:

- *The Kernel is hierarchical* – it refers to structure, in which the most abstract concepts are alphas, themselves having states; alphas are grouped in three-areas of concern (like layers in protocol stacks), which also serve to group activity spaces;
- *The Kernel is Actionable* – it refers to behavior, that can be associated with the Kernel, for instance it has activities;
- *The Kernel is extensible* – this is a property of modern modular software, it can be extended to allow additional functionality;
- *The Kernel is runnable* – indeed two parts of the Kernel book [1] are concerned with running iterations and software endeavors;

### B. The Bootstrap Principle for Software Theory

Based upon the above similarity of the SEMAT Kernel to software systems, we offer the following principle.

> **Bootstrap Principle** [for Software Theory]:
> A software theory should itself be runnable software.

The meaning of runnable is either that the theory may behave as a program which acts upon an input, or the theory is itself the input to another program.

This principle is motivated by the consequences one can infer from it and by analogy to previous software related efforts, as seen next.

### C. Desirable Consequences: Testability, Precision and Universality

Once we have a theory which is runnable in the above sense it can be actually tested – beyond gedanken experiments – say by making different hypotheses, applied to actual systems.

Running the theory makes the testing procedure amenable to quantitative criteria, thus more precise.

We can further test its universality, i.e. what are the limitations of the acceptable inputs.

### D. Bootstrap Analogies to Previous Software Efforts

The bootstrap idea has been often used in computers and software, e.g. in compilers. There are several programming languages and techniques that can be thought as runnable theories, many of them either related to AI (Artificial Intelligence), say LISP or PROLOG or to semantic representations, say using RDF or Ontologies.

For example LISP has been used both as a programming language and as a correctness proving language (see e.g. McCarthy and Talcott [7]), which can be said to embody a theory, being the set of criteria against which one can test correctness.

Within Software Engineering, similar ideas appeared in the context of software processes and software agents. For instance, Osterweill [8] stated that "Software Processes are Software Too".

## VI. THE ESSENCE: KERNEL ALPHAS AS A TOP LEVEL ONTOLOGY

As above stated, the Kernel is hierarchical. It is very natural to start the formulation of a theory matching the Kernel from its most abstract top level, viz. Kernel alphas. This is done in this section.

### A. Software Bottom-Up Levels of Abstraction

Let us have a closer look at the software levels of abstraction. A typical hierarchy describing an object-oriented software system has various levels of increasing abstraction in a bottom-up direction. The lowest level is machine code, the closest to a real or virtual machine. Next comes a higher level language, say Java. Going upwards another level, one finds a model, say in UML [9]. In recent proposals (see e.g. [10],[11]), one reaches an even higher level of abstraction – an ontology of concepts from which one can generate the UML classes.

### B. A Top-Level Ontology

A top-level ontology [12] is a compact representation of an application, as it has a relatively small number of concepts. The same is true of the Kernel alphas. Thus we formulate the theoretical matching Essence.

> **Essence Statement**:
> The SEMAT kernel alphas, the most abstract level of the kernel hierarchy, can be viewed as a top level ontology.

Viewing the kernel alphas as an ontology enables testing of the kernel properties, i.e. we enable checking the theory against empirical observations. This needs suitable tools, as illustrated in the following sub-sections.

### C. Kernel Alphas Ontology: Kinds of Things

We use the Protégé tool [13] developed at Stanford University, to edit the top level ontology, shown in Fig. 1. Essentially one has exactly the same graph as the kernel alphas in the Kernel book, with some noticeable additions. These additions are in response to questions motivated by the very fact that we are dealing with an ontology.

The topmost concept in the ontology is *thing*, equivalent to "object" in certain object-oriented languages. Then, we are driven to ask what kinds of concepts are in the top-level ontology.

From the bootstrap principle it follows that the things in the Kernel Alphas are *runnable*. A software-system is obviously runnable. A stakeholder or a member of a team, say a person is runnable in the sense that makes iterations with cards (cf. part II of the Kernel book [1]). Work is an activity, thus runnable. The reader may easily check by himself the other Alpha concepts.

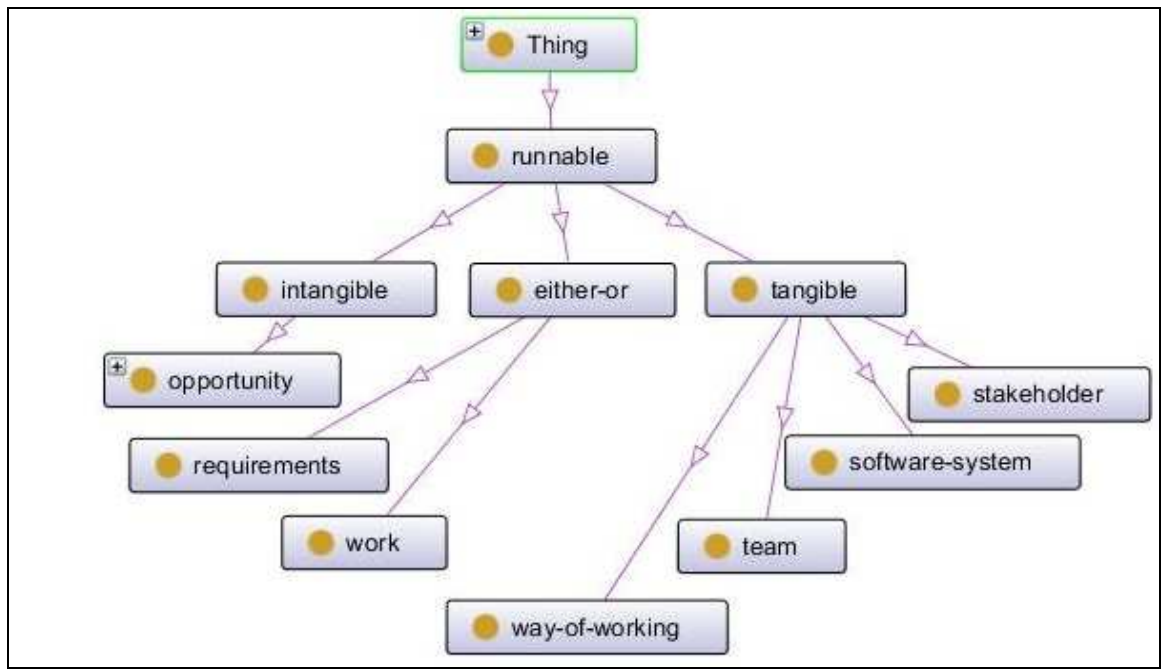

Figure 1. Top-level alphas' ontology as a schematic graph obtained from the Protégé tool. Each arrow means sub-classing – in the ontology direction convention – pointing to the sub-class. The topmost class is Thing. Note the specific additions – runnable, tangible, intangible and either-or – above the seven Kernel alphas. No other edges besides inheritance are shown. Explanations are given in the main text.

Some of the things are *tangible*, as they may contain some "hardware"[1], say the software system, stakeholder, team, way-of-working (i.e. a tool). In contrast, opportunity is *intangible*. The remaining things are classified as *either-or* as depending on the circumstances they may be either tangible or not. For instance, work may consist of motion of a physical robot arm or of some very abstract software calculation.

### D. Kernel Alphas Ontology: Completeness and Consistency

The properties of the Kernel Alphas can be tested by using an owl-xml representation supplemented by a set of axioms.
An owl-xml fragment on the "fulfill" property exported by the Protégé tool is shown in Fig. 2. A relevant traceability axiom states that "all requirements are fulfilled by modules of the software-system". This can be tested by a reasoner associated with the ontology tool and a convenient set of queries, say working on all the requirements for a given software-system.
Thus one can test the system for completeness and consistency, viz. absence of contradictions.

---

[1] Human biological tissue is taken as a kind of hardware.

```
...
<FunctionalObjectProperty>
        <ObjectProperty IRI="#fulfills"/>
</FunctionalObjectProperty>
<ObjectPropertyDomain>
        <ObjectProperty IRI="#fulfills"/>
        <Class IRI="#software-system"/>
</ObjectPropertyDomain>
<ObjectPropertyRange>
        <ObjectProperty IRI="#fulfills"/>
        <Class IRI="#requirements"/>
</ObjectPropertyRange>
```

Figure 2. Top-level alphas' ontology fragment, shown in the owl-xml representation, as exported by Protégé. It shows the "*fulfills*" object property, whose domain is the *software-system* class and whose range is the *requirements* class. This property is part of the original Kernel alphas, but not explicitly shown in Fig.1. IRI is Internationalized Resource Identifier, generalizing URI.

## VII. DISCUSSION

### A. Deeper Meaning of the Bootstrap Principle

The deeper meaning of the Bootstrap Principle is that runnable software guarantees that a software theory has all the nice required properties of a respectable theory.
It means:

- practical utility – by incorporating it into the Kernel for modern software development within a runnable environment;
- universality – by being implemented in a language equivalent to a Turing machine, thus applicable to any conceivable system;
- precision and measurability – by using quantitative criteria from the arsenal of software benchmarking tools;
- formality – by its very nature.
- 

One could also state that "Software is the only discipline in which entities can be their own theories". We are aware of the potential controversy of this statement, but we intentionally wish to trigger further discussion.

### B. Future Work

Planned future work includes extending the approach to specific added sub-alphas and other Kernel entities.
To reinforce the approach trustworthiness we intend to apply it to check actual realistic size software projects.

### C. Main Contribution

The main contribution of this paper is the Bootstrap Principle and its consequences, including the view of Kernel Alphas as an ontology.